\documentclass[11pt]{article}
\usepackage{amsmath}
\usepackage{amsfonts}
\usepackage{amssymb}
\usepackage{graphicx}

\def\bea{\begin{eqnarray}}
\def\eea{\end{eqnarray}}

\date{}

\begin{document}
\begin{center}
\LARGE { \bf  Cosmological New Massive Gravity and Galilean
Conformal Algebra in 2-dimensions  }
\end{center}
\begin{center}
{\bf M.R Setare\footnote{rezakord@ipm.ir} \\  V. Kamali\footnote{vkamali1362@gmail.com}}\\
{\ Department of Science, Payame Noor University, Bijar, Iran}
\\
\end{center}
\vskip 3cm
\begin{center}
{\bf{Abstract:} }
In the present paper we consider the realization
of $2-$dimensional Galilean conformal algebra ($GCA_2$) on the
boundary of cosmological new massive gravity. At first we
consider the contracted BTZ black hole solution. We obtain
entropy formula for the $GCA_2$ in term of contracted scaling
dimension $\Delta$ and central charge $C_1$. This entropy formula
 exactly matches with the non-relativistic limit of
Bekenstein-Hawking entropy of BTZ black hole. Then we extend our
study to the contracted warped $AdS_3$ black hole solution of
CNMG. We obtain the entropy of dual $GCA_2$ in terms of central
charges and finite temperatures, $T_1, T_2$. Again this
expression  coincides  with the non-relativistic limit of
Bekenstein-Hawking entropy formula of warped $AdS_3$ black hole.
\end{center}
\newpage
\section{Introduction}
Recently, there has been some interest in extending the AdS/CFT
correspondence to non-relativistic field theories \cite{a1},
\cite{b}. The Kaluza-Klein type framework for non-relativistic
symmetries, used in Refs. \cite{a1}, \cite{ b}, is basically
identical to the one introduced in \cite{d} (see also \cite{e}).
The study of a different non-relativistic limit was initiated in
\cite{f}, where the non-relativistic conformal symmetry obtained
by a parametric contraction of the relativistic conformal group.
Galilean conformal algebra (GCA) arises as a contraction
relativistic conformal algebras \cite{f}, \cite{c}, \cite{c1}, where in
$3+1$ space-time dimensions the Galilean conformal group is a
fifteen parameter group which contains the ten parameter Galilean
subgroup. Infinite dimensional  Galilean conformal group has been
reported  in \cite{c}, \cite{c1}, the generators of this group are :
~$L^n=-(n+1)t^n
x_i\partial_i-t^{n+1}\partial_t$,~$M^{n}_i=t^{n+1}\partial_i$
and~$J_{ij}^n=-t^n(x_i\partial_j-x_j\partial_i)$~ for an arbitrary
integer $n$, where  i and  j are specified by the spatial
directions. There is a finite dimensional subgroup of the
infinite dimensional Galilean conformal group which generated by
~($J_{ij}^0, L^{\pm 1}, L^0, M_i^{\pm 1}, M_i^0$). These
generators are obtained by contraction ($~t\rightarrow t, ~x_i
\rightarrow \epsilon  x_i, ~\epsilon \rightarrow 0,~ v_i \sim
\epsilon$ ) of the relativistic conformal generators. Recently
the authors of \cite{11} (see also \cite{bag1}) have shown that
the $GCA_2$ is the asymptotic symmetry of Cosmological
Topologically Massive Gravity (CTMG) in the non-relativistic
limit. They have obtained the central charges of $GCA_2,$ and
also a non-relativistic generalization of Cardy formula. In the
present paper we want to investigate similar problem for
Cosmological New
Massive Gravity (CNMG).\\
Recently, a new theory of massive gravity in three dimensions was
proposed by Bergshoeff, et. al \cite{2}. This theory, referred to
as "New Massive Gravity (NMG)", seems to offer a possibility for
formulating a consistent theory of quantum gravity in three
dimensions. NMG is defined by adding higher derivative term to
the EH term in action, with coupling $\frac{1}{m^2}$. If we add to
it the negative cosmological constant we can refer to this three
dimensional gravity as Cosmological New Massive Gravity (CNMG).
In this theory, the linearized excitations about the anti-de
Sitter vacuum describe a propagating massive graviton. One can
show that  CNMG  admits the BTZ black holes as
solutions \cite{2}, moreover in NMG regular warped black holes have been obtained by Clement \cite{a} (see also \cite{3}).\\
In this paper we propose the contracted BTZ and warped $AdS_3$
black hole solution of CNMG as a gravity dual of $2d$ GCA in the
context of the non-relativistic $AdS_3/CFT_2$ correspondence. The
rest of the paper is organized as: in section 2 we
give a brief review of 2d CFT and its contraction, the GCA parameters were realized in term of CFT parameters. In section
3, 4
we study GCA realization of on BTZ, and warped $AdS_3$ black hole solutions of NMG respectively, in
these sections GCA parameters were constructed in term of gravity
parameters and finally we obtained finite entropy in
non-relativistic limit. The last section is devoted to the
conclusion.
\section{Galilean Conformal Algebra in 2-Dimension}
Galilean conformal algebra in 2d can be obtained  from
contracting 2d conformal symmetry \cite{1}. 2$d$ conformal algebra
at the quantum level are described by two copy of Virasoro
algebra. In two dimensions space-time ($z=t+x$,
$\overline{z}=t-x$), the CFT generators
\begin{eqnarray}\label{1}
\mathcal{L}_n=-z^{n+1}\partial_{z},
~~~~~~~~~~~~~~~~~~~~~~~~\overline{\mathcal{L}}_n=-\overline{z}^{n+1}\partial_{\overline{z}}
\end{eqnarray}
obey the centrally extended Virasoro algebra
\begin{eqnarray}\label{2}
[\mathcal{L}_m,\mathcal{L}_n]=(m-n)\mathcal{L}_{m+n}+\frac{c_{R}}{12}m(m^2-1)\delta_{m+n,0}\\
\nonumber
[\overline{\mathcal{L}}_m,\overline{\mathcal{L}}_n]=(m-n)\overline{\mathcal{L}}_{m+n}+\frac{c_{L}}{12}m(m^2-1)\delta_{m+n,0}
\end{eqnarray}
By taking the non-relativistic limit ($t\rightarrow t$,
$x\rightarrow\epsilon x$ with $\epsilon\rightarrow 0$),  the GCA
generators $L_{n}$ and $M_n$ are constructed from Virasoro
generators by
\begin{eqnarray}\label{3}
L_n=\lim_{\epsilon\rightarrow
0}(\mathcal{L}_n+\overline{\mathcal{L}}_n)~~~~~~M_{n}=\lim_{\epsilon
\rightarrow 0 }\epsilon (\mathcal{L}_n-\overline{\mathcal{L}}_n)
\end{eqnarray}
From Eqs.(\ref{2}) and (\ref{3}), one obtains centrally extended 2d GCA
\begin{eqnarray}\label{4}
[L_m,L_n]=(m-n)L_{m+n}+C_1m(m^2-1)\delta_{m+n,0}~~\\
\nonumber [L_m,M_n]=(m-n)M_{m+n}+C_2m(m^2-1)\delta_{m+n,0}\\
\nonumber [M_n,M_m]=0~~~~~~~~~~~~~~~~~~~~~~~~~~~~~~~~~~~~~~~~~~~~~~
\end{eqnarray}
Note that $[M_n,M_m]$ cannot have any central extension. The GCA
central charges ($C_1$, $C_2$) are related to CFT central charges
($c_L$, $c_R$) as:
\begin{eqnarray}\label{5}
C_1=\lim_{\epsilon\rightarrow
0}\frac{c_L+c_R}{12}~~~~~~C_2=\lim_{\epsilon\rightarrow
0}(\epsilon\frac{c_L-c_R}{12})
\end{eqnarray}
Similarly, rapidity $\xi$ and scaling dimensions $\Delta$, which
are the eigenvalues of $M_0$ and $L_0$ respectively, are given by
\begin{eqnarray}\label{6}
\Delta=\lim_{\epsilon \rightarrow
0}(h+\overline{h})~~~~~~~~~\xi=\lim_{\epsilon\rightarrow
0}\epsilon (h-\overline{h})
\end{eqnarray}
where $h$ and $\overline{h}$ are eigenvalues of $L_0$ and
$\overline{L}_0$ respectively. So the 2$d$ GCA was obtained by the
non-relativistic limit of the Virasoro $CFT_2$.
\section{GCA realization of on BTZ black hole solution of NMG }In this section we would like
to propose that the contracted BTZ black hole solution of  three
dimensional new massive gravity (NMG) is gravity dual of 2$d$ GCA
in the context of $AdS/CFT$ correspondence. It is notable that the
NMG (as a gravity dual) has to yield finite parameters ($\Delta,
\xi, C_1, C_2$ and entropy $S_{GCA}$) for GCA. The action of the
cosmological new massive gravity in three dimension is \cite{2}
\begin{eqnarray}\label{7}
S[g_{\mu\nu}]=\frac{1}{16\pi G}\int \sqrt{-g}(R-2\lambda
+\frac{1}{m^2}\mathcal{L}_{NMG})
\end{eqnarray}
where the NMG term is
\begin{eqnarray}\label{8}
\mathcal{L}_{NMG}=R_{\mu\nu}R^{\mu\nu}-\frac{3}{8}R^2
\end{eqnarray}
The Einstein equation of motion of this action is
\begin{eqnarray}\label{a}
G_{\mu\nu}+\lambda m^2 g_{\mu\nu}-\frac{1}{2m^2}K_{\mu\nu}=0
\end{eqnarray}
where
\begin{eqnarray}\label{b}
K_{\mu\nu}=-\frac{1}{2}\nabla^2Rg_{\mu\nu}-\frac{1}{2}\nabla_{\mu}\nabla_{\nu}R+2\nabla^2R_{\mu\nu}+4R_{\mu\alpha\nu\beta}R^{\alpha\beta}\\
\nonumber
-\frac{3}{2}RR_{\mu\nu}-R_{\alpha\beta}R^{\alpha\beta}g_{\mu\nu}+\frac{3}{2}R^2g_{\mu\nu}~~~~~~~~~~~~~~~~~~~~~~~~
\end{eqnarray}
where $g^{\mu\nu}K_{\mu\nu}=\mathcal{L}_{NMG}$. The parameter
$\lambda$ is dimensionless and characterizes the cosmological
constant term, while $m$ has the dimension of mass and provides
the coupling to the NMG term. The solution of BTZ black hole is
given by
\begin{eqnarray}\label{9}
ds^2=(-f(r)+\frac{16G^2J^2}{r^2})dt^2+\frac{dr^2}{f(r)}+r^2d\varphi^2+8GJdtd\varphi
\end{eqnarray}
where
\begin{eqnarray}\label{10}
f(r)=(\frac{r^2}{l^2}-8GM+\frac{16G^2J^2}{r^2})~~~l^{-2}=2m^2(-1\pm\sqrt{1-\frac{\lambda}{m^2}})
\end{eqnarray}
The parameters $M$ and J correspond to the mass and angular
momentum in the case without the new massive gravity term, but
their definitions in the case with NMG term are \cite{a}
\begin{eqnarray}\label{11}
M_1=(1-\frac{1}{2m^2l^2})M~~~~~~~~~~J_1=(1-\frac{1}{2m^2l^2})J
\end{eqnarray}
Due to NMG term, the Bekenstein-Hawking entropy is renormalized
by the same factor $(1-\frac{1}{2m^2l^2})$
\begin{eqnarray}\label{12}
S_{BH}=\frac{\pi r_h}{2G}(1-\frac{1}{2m^2l^2}),
\end{eqnarray}
here $r_h=\sqrt{2Gl(lM+J)}+\sqrt{2Gl(lM-J)}$. In the paper \cite{sun} Liu and Sun have analyzed the
Brown-Henneaux boundary condition for the NMG. They have calculated the conserved charges corresponding
to the generators of the asymptotical symmetry under the Brown-Henneaux boundary condition (see the appendix).
Then they have obtained the central charges of the Virasoro algebra as \cite{sun},\cite{31}
\begin{eqnarray}\label{13}
c_L=c_R=\frac{3l}{2G}(1-\frac{1}{2m^2l^2})
\end{eqnarray}
For the BTZ black hole solution in NMG, $h$ and $\overline{h}$
are calculated as
\begin{eqnarray}\label{14}
h=\frac{1}{2}(lM_1+J_1)+\frac{c_L}{24}~~~~~~~~\overline{h}=\frac{1}{2}(l
M_1-J_1)+\frac{c_R}{24}
\end{eqnarray}
then the microscopic entropy is expressed by Cardy formula
\begin{eqnarray}\label{15}
S_{CFT}=2\pi(\sqrt{\frac{c_L
h}{6}}+\sqrt{\frac{c_R\overline{h}}{6}})=\frac{\pi
r_h}{2G}(1-\frac{1}{2m^2l^2})
\end{eqnarray}
which  agrees with renormalized Bekenstein-Hawking entropy
formula Eq.(\ref{12}). Now let us consider non-relativistic limit
in three dimensional gravity.
\begin{eqnarray}\label{16}
t\rightarrow t~~~~~~~r\rightarrow
r~~~~~~~\varphi\rightarrow\epsilon \varphi
\end{eqnarray}
Accordingly, the parameters $M$ and $J$ in the BTZ solution must
scale like
\begin{eqnarray}\label{17}
M\rightarrow M~~~~~~~ J\rightarrow \epsilon J
\end{eqnarray}
As have been discussed in \cite{11}, the black hole metric (\ref{9}) degenerates and looks singular in the Galilean limits (\ref{16}), (\ref{17}). This is similar to the usual Newtonian approximation $c\rightarrow\infty$. This situation in the bulk gravity is describe by the Newton-Cartan-like geometry for the geometry  with the $AdS_2$ base \cite{c,11}.\\ Since the Virasoro generators corresponds to $\mathcal{L}^{\pm}_n=i\xi_n^{R,L}$, from the gravity side using Eqs. (A.3),(\ref{16}) and (\ref{3}) one can define the generators $L_n$ and $M_{n}$ as
\begin{eqnarray}\label{}
L_n=e^{in\tau}[(1-2e^{-2\rho}n^2)\partial_{\tau}+in\phi(1+2e^{-2\rho}n^2)\partial_{\phi}-in\partial_{\rho}]\\
\nonumber
M_n=e^{in\tau}(1+2e^{-2\rho}n^2)\partial_{\phi}~~~~~~~~~~~~~~~~~~~~~~~~~~~~~~~~~~~~~~~~
\end{eqnarray}
these generators satisfy the center-less version of GCA algebra (\ref{4}).
From Eqs.(\ref{5}) and (\ref{13}) the GCA central charges $C_1$
and $C_2$ in this case are
\begin{eqnarray}\label{18}
C_1=\frac{l}{4G}(1-\frac{1}{2m^2l^2})~~~~~~C_2=0
\end{eqnarray}
So, one of the GCA central charges vanishes because of the parity
invariance of the NMG.\\
Similarly, from Eqs.(\ref{6}) and
(\ref{14}), scaling dimensions $\Delta$ and rapidity $\xi$, which
are the eigenvalue of $L_0$ and $M_0$ are given by
\begin{eqnarray}\label{19}
\Delta=\lim_{\epsilon\rightarrow 0} (l
M_1+\frac{c_L+c_R}{24})=(1-\frac{1}{2m^2l^2})lM+\frac{C_1}{2}\\
\nonumber \xi=\lim_{\epsilon\rightarrow
0}\epsilon(J+\frac{c_L-c_R}{24})=0~~~~~~~~~~~~~~~~~~~~~~~~~~~~
\end{eqnarray}
When $M$ are large enough, the last term $\frac{C_1}{2}$ can be
neglected.\\
Now we want to obtain the entropy of the GCA. The
scaling limit ($M\rightarrow M$, $J\rightarrow \epsilon J$)
requires that the event horizon of the BTZ black hole should scale
~$r_h\rightarrow 2l\sqrt{2GM}$, so the black hole  entropy
(\ref{12}) is given by
\begin{eqnarray}\label{20}
\lim_{\epsilon \rightarrow 0 }S_{BH}=\lim_{\epsilon \rightarrow 0
}S_{CFT}=\pi(1-\frac{1}{2m^2l^2})\sqrt{\frac{2l^2M}{G}}
\end{eqnarray}
From the expressions of the central charge $C_{1}$ (\ref{18}) and
scaling dimension $\Delta$ (\ref{19}), we can rewrite the entropy
as
\begin{eqnarray}\label{20}
S_{GCA}=\pi\sqrt{8\Delta C_1 }
\end{eqnarray}
This expression is the entropy for the GCA in two dimension.
\section{GCA realization on warped $AdS_3$ black hole solution of NMG}
In this section we consider the warped black holes in NMG. The
metric of warped NMG can be written as \cite{a, 3}
\begin{eqnarray}\label{21}
\frac{ds^2}{l^2}=dt^2+\frac{dr^2}{\zeta^2\eta^2(r-r_+)(r-r_-)}+\mid\zeta\mid(r+\eta\sqrt{r_+r_-})dtd\varphi\\
\nonumber
+\frac{\zeta^2}{4}r((1-\eta^2)r+\eta^2(r_++r_-)+2\eta\sqrt{r_+r_-})d\varphi^2~~~~~~~~~~~~~~~~~~
\end{eqnarray}
The ranges of the coordinates are $t\in(-\infty, +\infty)$,
$r\in[0,+\infty)$ and $\varphi\in[0, 2\pi]$. There are two
horizons which are located at $r_+$ and $r_-$. The parameters
$\eta$, $\zeta$ are given by
\begin{eqnarray}\label{23}
\eta^2=\frac{21}{4}+\frac{2m^2l^2}{\zeta^2}~~~~~~~~~~\zeta^2=\frac{4m^2l^2}{21}(-6\pm\sqrt{3(5-7\lambda)})
\end{eqnarray}
The Bekemstein-Hawking entropy in this case is
\begin{eqnarray}\label{24}
S_{BH}=\frac{\pi l\zeta^3}{2m^2G}(r_++\eta\sqrt{r_+r_-})
\end{eqnarray}
The microscopic entropy of the dual CFT can be computed by the
Cardy formula which matches with the black hole
Bekenstein-Hawking entropy \cite{3}
\begin{eqnarray}\label{25}
S_{CFT}=\frac{\pi^2l}{3}(c_LT_L+c_RT_R)
\end{eqnarray}
The left and right moving temperatures introduced in \cite{3}
\begin{eqnarray}\label{26}
T_L=\frac{\eta^2\zeta^2}{8\pi
l}(r_++r_-+2\eta\sqrt{r_+r_-})~~~~~~~T_R=\frac{\eta^2\zeta^2}{8\pi
l }(r_+-r_-)
\end{eqnarray}
By the calculation such as we review briefly in appendix, one can obtain the central charges of warped
$AdS_3$ black hole solution of NMG as (see also \cite{3})
\begin{eqnarray}\label{27}
c_L=c_R=\frac{6l\zeta}{G\eta^2m^2}
\end{eqnarray}
From the above central charges and temperatures (\ref{26}) we have
\begin{eqnarray}\label{28}
S_{CFT}=\frac{\pi l\zeta^3}{2m^2G}(r_++\eta\sqrt{r_+r_-})
\end{eqnarray}
which agrees precisely with the gravity result presented in
\cite{3}. Now we consider non-relativistic limit in three
dimensional new massive  gravity.
\begin{eqnarray}\label{288}
t\rightarrow t~~~~~~~r\rightarrow
r~~~~~~~\varphi\rightarrow\epsilon \varphi
\end{eqnarray}
GCA central charges $C_1$ and $C_2$ are defined in
terms of CFT central charges (\ref{5})
\begin{eqnarray}\label{}
C_1=\frac{l\zeta}{G\eta^2m^2}~~~~~~C_2=0
\end{eqnarray}
  CFT entropy (\ref{25})
with the limit ($\epsilon\rightarrow 0$) is converted to Galilean
conformal entropy.
\begin{eqnarray}\label{29}
S_{GCA}=\lim_{\epsilon\rightarrow
0}(\frac{\pi^2}{3}[6C_1(T_L+T_R)+6C_2(\frac{T_L-T_R}{\epsilon})])
\end{eqnarray}
We define Galilean conformal  temperatures as following
\begin{eqnarray}\label{30}
T_1=\lim_{\epsilon\rightarrow 0}
6(T_L+T_R)~~~~~~~~T_2=\lim_{\epsilon\rightarrow
0}6\frac{T_L-T_R}{\epsilon}
\end{eqnarray}
the GCA entropy is
\begin{eqnarray}\label{31}
S_{GCA}=\frac{\pi^2}3({C_1T_{1}+C_{2}T_{2}})~
\end{eqnarray}
To make the GCA entropy finite, $T_L+T_R\sim O(1)$ and
$T_L-T_R\sim O(\epsilon)$
\begin{eqnarray}\label{32}
T_L+T_R=\frac{\eta^2\zeta^2}{4\pi l}(r_++\eta\sqrt{r_+r_-})
~~~~~~~~~~ T_L-T_R=\frac{\eta^2\zeta^2}{4\pi
l}(r_-+\eta\sqrt{r_+r_-})
\end{eqnarray}
The parameters $r_+$ and $r_-$ in the non-relativistic warped NMG
must scale like
\begin{eqnarray}\label{33}
r_+\rightarrow r_+~~~~~~~~~~~~r_-\rightarrow \epsilon^2r_-
\end{eqnarray}
Then, $T_L+T_R=\frac{\eta^2\zeta^2}{4\pi l}r_+\sim O(1)$ and
$T_L-T_R=\frac{\eta^2\zeta^2}{4\pi l}\epsilon\sim O(\epsilon)$,
and the GCA entropy is finite
\begin{eqnarray}\label{34}
S_{GCA}=\frac{\pi l\zeta^3}{2m^2G}r_+
\end{eqnarray}
which  agrees  with Bekenstein-Hawking entropy of warped CNMG in
non-relativistic limit (\ref{33}).
\section{Conclusion} In the present paper we have considered the
BTZ and Warped $AdS_3$ black hole solution of CNMG in the
non-relativistic limit. The BTZ solution of CNMG has been
obtained from a BTZ solution of pure Einstein gravity only by
redefinition of mass and angular momentum parameter of metric, as
Eq.(\ref{11}), \cite{a}. We have shown that the contracted BTZ
solution has a dual description in term of a $2-$dimensional GCA.
We have obtained the central charges $C_1$, $C_2$ and also the
scaling dimension $\Delta$ and rapidity $\xi$ as equations
(\ref{18}), (\ref{19}) respectively. Using these results and
scaling limits ($M\rightarrow M, J\rightarrow \epsilon J$), we
have obtained the entropy of $GCA_2$ by Eq.(\ref{20}), this
expression  exactly agrees  with the non-relativistic limit of
Bekenstein-Hawking entropy of the black hole. After that we
studied the warped $AdS_3$ black hole solution in
non-relativistic limit. Using the central charges $C_1$, $C_2$
for $GCA_2$, we obtained the entropy of non-relativistic
$CFT_2$,  as Eq.(\ref{29}). Then we have defined Galilean
conformal temperatures by Eq.(\ref{30}). Using these results we
have obtained final form of $GCA$ entropy by Eq.(\ref{34}), which
is exactly the non-relativistic limit of entropy formula
(\ref{24}).
\section{Appendix}
Here we give a brief review of Brown-Henneaux boundary condition for the NMG (to see more
about it refer to \cite{sun}). The NMG have an $AdS_3$ solution\\
\\
~~~~~~~~~~~~~~$ds^2=\overline{g}_{\mu\nu}dx^{\mu}dx^{\nu}=L^2(-\cosh^2\rho~ d\tau^2+\sinh^2\rho ~d\phi^2+d\rho^2)~~~~~~(A.1)$\\
\\
by expanding $g_{\mu\nu}=\overline{g}_{\mu\nu}+h_{\mu\nu}$ around $AdS_3$, one could obtain the equation of motion for the linearized exitations $h_{\mu\nu}$. The Brown-Henneaux boundary condition for the linearized gravitational exitations in asymptotical $AdS_3$ spacetime in the global coordinate system can be written as \\

~~~~~$\left(\begin{array}{ccccc}
 h_{++}=\mathcal{O}(1)&~ h_{+-}=\mathcal{O}(1)&~ h_{+\rho}=\mathcal{O}(e^{-2\rho}) \\
 h_{-+}=h_{+-}&~h_{--}=\mathcal{O}(1)&~h_{-\rho}=\mathcal{O}(e^{-2\rho})\\
 h_{\rho +}=h_{+\rho}&~h_{\rho -}=h_{-\rho}&~h_{\rho\rho}=\mathcal{O}(e^{-2\rho})\\
  \end{array}\right)\label{h}~~~~~~~(A.2)$
 \\
 \\
 The corresponding asymptotic Killing vectors are \\
 \\
 $\xi=\xi^{+}\partial_{+}+\xi^{-}\partial_{-}+\xi^{\rho}\partial_{\rho}=[\epsilon^{+}(\tau^{+})+2e^{-2\rho}\partial_{-}^2\epsilon^{-}(\tau^{-})
 +\mathcal{O}(e^{-4\rho})]\partial_{+}\\
 \\
 +[\epsilon^{-}(\tau^{-})+ 2e^{-2\rho}\partial_{+}^{2}\epsilon^{+}(\tau^{+})+\mathcal{O}(e^{-4\rho})]\partial_{-}\\
 \\
 -\frac{1}{2}[\partial_{+}\epsilon^{+}(\tau^{+})+\partial_{-}\epsilon^{-}(\tau^{-})+\mathcal{O}(e^{-2\rho})]\partial_{\rho}
 ~~~~~~~~~~~~~~~~~~~~~~~~~~~~~~~(A.3)$\\
 \\
 where $\tau^{\pm}=\tau\pm\phi,~ \partial_{\pm}=\frac{1}{2}(\partial_{\tau}\pm\partial_{\phi})$.
 Since $\phi$ is periodic, one could choose the basis $\xi_m^{+}=e^{im\tau^{+}}$ and $\xi^{-}_n=e^{in\tau^{-}}$ and denote
 the corresponding Killing vectors as $\xi_m^{L}$ and $\xi_n^{R}$. The algebra structure of these vectors is\\
 \\
 $i[\xi_m^L,\xi_n^L]=(m-n)\xi^{L}_{m+n},~~~~~i[\xi_m^R,\xi_n^R]=(m-n)\xi^{R}_{m+n}$ \\
 \\
 $[\xi_m^L,\xi_n^R]=0~~~~~~~~~~~~~~~~~~~~~~~~~~~~~~~~~~~~~~~~~~~~~~~~~~~~~~~~~~~~~~~~~~~~~(A.4)$\\
 \\
 So these asymptotic Killing vectors give two copies of Virasoro algebra. Then by calculating the conserved charges, the authors of \cite{sun} have shown that the left moving and right moving conserved charges fulfil two copies of Virasoro algebra with central charges (\ref{13}).

\end{document}